\documentclass[preprint2]{aastex}
\tightenlines

\def\teff {$T_{\rm eff}$}

\begin{document}

\title{The Predicted Signature of Neutrino Emission in Observations of Pulsating
Pre-White Dwarf Stars}

\author{ 
M.~S.~O'Brien\altaffilmark{1} 
\and
S.~D.~Kawaler\altaffilmark{2} }

\altaffiltext{1}{Space Telescope Science Institute, 3700 San Martin Drive,
Baltimore, MD 21218; obrien@stsci.edu.}
\altaffiltext{2}{Department of Physics and Astronomy, Iowa State University,
Ames, IA 50011.}

\begin{abstract}

Pre-white dwarf (PWD) evolution can be driven by energy losses from neutrino
interactions in the core.  Unlike solar neutrinos, these are not the
by-product of nuclear fusion, but instead result from electron scattering
processes in the hot, dense regions of the PWD core.

We show that the observed rate of period change in cool PWD pulsators 
will constrain neutrino emission in their cores, and we identify appropriate 
targets for future observation.  Such a measurement will tell us
whether the theories of lepton interactions correctly describe the production 
rates and therefore neutrino cooling of PWD evolution.  This would represent the
first test of standard lepton theory in dense plasma.

\end{abstract}


\section{Introduction}

In general, stars are too remote---and observables too few---to make 
them practical experimental physics test-beds:  our data are spent in simply 
describing the dimensions of the objects under study.  In many cases we must 
extrapolate experimental data over many orders of magnitude, or resort 
to untested calculations from first principles, to reach the regions 
of phase space that apply to stellar interiors.

If we hope to overcome these problems and pursue ``experimental'' astrophysics, 
we can either attempt to increase the number of observables or find simpler 
stars.  As first realized by Mestel (1952), the evolution of white
dwarfs and pre-white dwarfs (PWDs) is primarily a simple cooling problem.  In 
general, our growing understanding of white dwarf interiors and evolution has 
paralleled advances in the theory of dense plasmas, with the recognition of important
influences like electron degeneracy (Chandrasekar 1939), Coulomb interactions 
(Salpeter 1961), crystallization (Kirzhnitz 1960; Abrikosov 1960; Salpeter 1961; 
Stevenson 1980) and neutrino cooling effects (Chin, Chiu, \& Stothers 1966; Winget, 
Hansen, \& Van Horn 1983; Kawaler, Hansen, \& Winget 1985). Iben \& Tutukov 
(1984) summarize the various mechanisms which dominate
white dwarf evolution from the planetary nebula nucleus (PNN) stage to 
the coolest white dwarfs.

On the observational side, the discovery of white dwarf pulsation in the 1960s,
and pre-white dwarf pulsation in the 1970s, greatly increased the observable 
parameters available for comparison with theoretical models.  These are
short period, multiperiodic, $g$-mode variables, showing anywhere from a few to
over a hundred separate periodicities on timescales of 100-3000~s.  The pulsating
PWD stars are divided into two classes:  the planetary nebula nucleus variables
(PNNV stars), and the slightly more evolved GW~Virginis (or simply GW~Vir stars)
which lack observed nebulae.  With high surface gravities (log~$g \sim 6-7.5$),
and effective temperatures between 80,000~K and 170,000~K, they occupy a region 
of the H-R diagram between the high-\teff~end of the PNN branch and the top of 
the white dwarf cooling track.  There are eight known PNNV stars, and four 
GW~Vir stars (Ciardullo \& Bond 1996).

The evolutionary timescale of PWD stars is of order $10^{6}$ years.  During this
short transition from PNN star to hot white dwarf, stellar radius and photon
luminosity decrease by one and three orders of magnitude, respectively.
High core density and temperature allow electron scattering processes to produce
a large neutrino flux which remains roughly constant during this time.
As photon luminosity plummets, neutrinos contribute an increasing fraction of the total
energy losses.  Neutrino emission eventually comes to dominate the 
overall evolution of the star.

Unlike photon energy, which must diffuse relatively slowly through
the entire star before emerging into space, neutrinos created
near the center of the PWD escape directly.
This neutrino luminosity cools the center of the star, maintaining a
temperature inversion similar to that within stars at the tip of the red
giant branch.  Calculations of
the relevant reaction rates were performed initially by Beaudet,
Petrosian, \& Salpeter (1967) based on the theory of weak interactions
proposed by Feynman \& Gell-Mann (1958).  Later, Dicus (1972) and Dicus
et al.~(1976) recalculated these rates in the unified electroweak theory
of Weinberg and Salam (Weinberg 1967, Salam 1968).  All of these
calculations are theoretical, however.  We have no direct experimental
or observational confirmation of neutrino production rates under
conditions appropriate to PWD interiors.

The cooling of a GW~Vir interior tends to increase the periods of 
each given pulsation mode.  Their high luminosity (log~$L \sim 0 - 3$) 
means they cool much more rapidly than cooler white dwarf variables.
GW~Vir period changes are therefore expected to be more 
rapid also.  Winget, Hansen \& Van Horn (1983) show that the 
$e$-folding time for period changes in GW~Vir stars should be 
of the same order as the evolutionary timescale---$10^{6}$ years; such rapid changes 
are measurable in $1-3$ years time.  This is an exciting prospect: to measure directly,
on human timescales, the rate of evolution of a star, and specifically 
to place strict constraints on the mechanisms which regulate the evolution
of a stellar interior.  Over 30 years ago, Chin,
Chiu, and Stothers (1966) predicted that at some point in PWD evolution 
neutrino losses should dominate all other cooling processes.  Asteroseismological
analysis can tell us which stars these are, and then measurement of
period changes can tell us if our neutrino physics is right.

Such a test has implications far beyond the study of
PWD evolution.  For instance, one of the fundamental 
questions of stellar astrophysics is the length of time stars spend on 
the main sequence.
Answering this question requires precise knowledge of the p-p and CNO 
nuclear reaction rates.  Currently, the best laboratory for measuring 
these rates is our own Sun, since terrestrial labs cannot in general 
reproduce the conditions of the stellar interior.  However, models 
which successfully reproduce the known structure of the Sun predict a 
neutrino flux two to three times that measured by earthly detectors 
(Bahcall \& Pinsonneault 1996, and references therein).  For a long 
time, it was thought the problem might reside 
in our incomplete knowledge of conditions in the solar interior.  Recently, 
helioseismology projects such as the Global Oscillation Network Group 
(GONG) have resulted in the measurement of millions of solar pulsation 
frequencies (Harvey et al.~1996). With so many parameters to constrain model 
properties, the possibility that the solar neutrino problem can be solved
through variations in the thermodynamics or mechanics seems to be excluded 
(Bahcall \& Pinsonneault 1996).  The problem, then, almost certainly lies 
with the way we handle the nuclear physics.  

Under the most intense scrutiny is the standard theory of lepton interactions.  
Our calculations of neutrino emission from PWDs are based on this
same theory.  In PWDs, however, the energy loss rate due to
neutrinos is thousands of times greater than in the Sun.  Measurement of
the effects of neutrino interactions in PWDs would afford a
critical independent test not only of the standard lepton theory but also
of non-standard theories brought forward to solve the solar neutrino
problem.

To explore this possibility, we calculated PWD evolutionary 
tracks using different neutrino production rates.  In the next section 
we describe the calculation of those rates and summarize the basic
interactions that lead to neutrino emission in PWD interiors.
Section~3 describes PWD sequences with varied
neutrino production rates and examines effects on measurable quantities 
such as $T_{\rm eff}$, surface gravity, and rate of period change.  
Finally, in \S~4 we discuss prospects for placing observational
constraints on neutrino physics, and we identify appropriate targets 
for future observation.

\section{Neutrino Cooling in Pre-White Dwarf Interiors}

Unlike the solar neutrino flux, neutrino emission in PWDs is not a 
by-product of nuclear fusion.  Instead, the 
density and temperature in their cores are high enough 
(log~$\rho_{\rm c} \sim$~6--7, log~$T_{\rm c} \sim$~7--8) to produce 
neutrinos directly through several different scattering processes.  The 
two most important processes are {\it neutrino bremmstrahlung} and 
{\it plasmon} excitation.   Neutrino bremmstrahlung is much like the 
normal bremmstrahlung process familiar to astrophysicists, in which
high-energy electrons scatter off nuclei, emitting X-rays.  At the 
high density and temperature of PWD interiors, however, 
neutrinos can be produced instead.  
%
% New, in response to comments by the referee:
%
These same conditions support the existence of thermally excited 
photons within the plasma, analogous to phonons propagating within
a metal lattice.  These ``plasmons'' have a finite lifetime and decay 
to form a neutrino and antineutrino.\footnote{Actually there are two
types of plasmons.  The process described here is that of the 
{\it transverse} plasmon.  The other, {\it longitudinal} plasmon 
corresponds to an oscillation in the electron gas similar to a
sound wave, but is usually less important as a neutrino source in 
hot white dwarfs (Itoh et al.~1992).}
%
% End New 
%
%emissionderives from the dispersion relation of a photon in an electron 
%gas, given by
%\begin{equation}
%(\hbar\omega)^{2} = (\hbar\omega_{\rm 0})^{2} + (\hbar kc)^{2}
%\end{equation}
%where $\omega$ is the photon angular frequency, $k$ is the wave number of
%the photon, and $\omega_{\rm 0}$ is the so-called plasma frequency.
%The plasma frequency depends on the electron temperature and density,
%and vanishes in free space.  A photon obeying Equation~(1) behaves
%like a particle with an effective mass of $\hbar\omega_{\rm 0}/c^{2}$, 
%and is therefore called a {\it plasmon}.  Their effective mass 
%means that (unlike photons in free space) plasmons can decay directly 
%into electron-positron pairs which then annihilate into electron neutrinos 
%and antineutrinos.

The possible relevance of the plasma process to stellar astrophysics was 
first pointed out by Adams, Ruderman, \& Woo (1963), who subsequently 
calculated rates based on the theory of Feynman \& Gell-Mann (1958).
Beaudet, Petrosian \& Salpeter (1967) were the first to incorporate them
into stellar evolution calculations.
Later, Dicus (1972) recalculated the rates of various neutrino processes
in the unified electro-weak theory of Weinberg and Salam (Weinberg 1967, 
Salam 1968).  

The rates used in our stellar evolution code, ISUEVO, derive from updated 
calculations by Itoh et al.~(1996), and include the plasmon, 
bremmstrahlung, and several less important neutrino production processes.
The evolution code ISUEVO (Dehner 1996; see also Dehner \&
Kawaler 1995) is optimized for the construction of PWD and white dwarf
models.  The models used in this investigation are based on the evolution
of a $3~M_{\odot}$ model from the Zero Age Main Sequence through the thermally 
pulsing AGB phase.  After reaching a stable thermally pulsing stage 
(about 15 thermal pulses), mass loss was invoked until the model evolved to 
high temperatures.  This model (representing a PNN) had a final mass of 
$\sim 0.6~M_{\odot}$, and a helium-rich outer layer.  Additional details 
concerning the construction of this evolution sequence (and others of different
mass, discussed in \S~3, below) can be found in O'Brien (1998).

To study the direct effects of neutrino losses on PWD
evolution, we introduced artificially altered rates well before the 
evolving models reached the PWD track.  If we simply changed 
the rates beginning at the hot end of the PWD sequence, 
the thermal structure of each model would take several thermal 
timescales to relax to a new equilibrium configuration
based on the new rates.  Unfortunately, this relaxation time is of
the same order as the PWD cooling time, and so only the
cool end of the sequence would see the full effects of the new rates
on their evolutionary timescales.  Therefore, the enhanced and diminished
rates described in the next section were introduced into evolutionary
calculations beginning at the base of the AGB.  The resulting thermal
structure of the initial PWD ``seed'' models was then already
consistent with the neutrino rates used during the prior evolution
that produced them.

\section{Pre-White Dwarf Sequences with Different Neutrino Rates}

Starting with the PWD seed models above, we
evolved the models from high $L$ and $T_{\rm eff}$
down toward the white dwarf cooling track.  Three sequences were
calculated.  The base sequence used the normal neutrino production rates.
Another sequence used rates diminished by a factor of three (at 
any given $\rho$ and $T$ in the stellar interior) over the normal rates, 
while the third sequence used rates enhanced by a factor of three.  
This trio spans nearly one order of magnitude in neutrino production.

The resulting $0.6\,M_{\odot}$ evolutionary sequences are shown in
Figure~1, from $T_{\rm eff} \sim 170,000\,K$---equivalent to the hottest
PWDs known---down to about $35,000\,K$.
Luminosity decreases by almost four orders of magnitude in approximately
five million years.  The GW~Vir instability strip occupies the left half
of the figure, above $T_{\rm eff} \sim 80,000\,K$ (log~$T_{\rm eff} = 4.9$), 
a temperature reached by the PWD models in only 500,000 years.

\begin{figure}[h]
\includegraphics[scale=0.29,angle=-90]{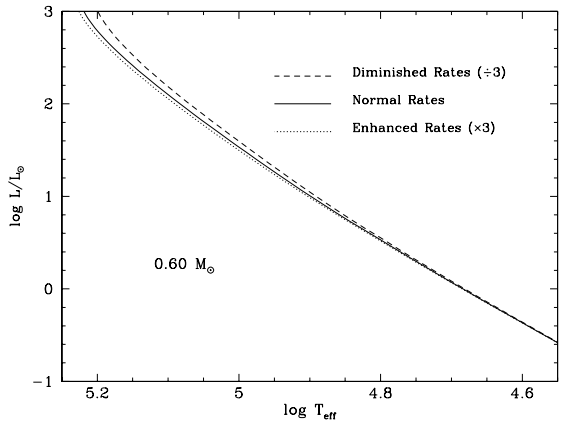}
\caption{Evolutionary tracks for three $0.6~M_{\odot}$ model
sequences with different input neutrino production rates.
The upper and lower tracks were calculated with rates one-third and
three times the normal rates (middle track), respectively.}
\end{figure}

The most striking aspect of Figure~1 is the similarity of the tracks:
changing the neutrino rates seems to have little effect on the
luminosity at a given $T_{\rm eff}$ at {\it any} point
in PWD evolution, despite the importance of neutrino
losses as a cooling mechanism over much of this range.  In
Figure~2, we find that, for all three sequences, neutrino losses
are the {\it primary} cooling mechanism over the approximate range
$100,000\,K < T_{\rm eff} < 30,000\,K$.  Plasmon reactions dominate over 
the bremmstrahlung process for 0.6~$M_{\odot}$ models at all stages of 
PWD evolution, as shown in Figure~3.

The ratio $L_{\nu}/L_{\gamma}$ also increases with stellar mass.
In the \teff~range 80,000--100,000~K, $L_{\nu}/L_{\gamma}$ for a 
0.66~$M_{\odot}$ model sequence is nearly 30\% higher than for a 
0.60~$M_{\odot}$ sequence.

Figures~1 and 2 show that the differences in
$L$ and $T_{\rm eff}$ are smallest when the neutrinos are important.
This is because the primary {\it structural} effect of changing the
neutrino rates is on the radius of the models (Figure~4), causing the
tracks to assume a position in the $L$--$T_{\rm eff}$ plane normally
occupied by models of slightly higher mass (for enhanced rates) or lower
mass (for diminished rates).  However, at lower temperatures electron 
degeneracy becomes increasingly important as a mechanical
support against gravity (and thus in determining the final stellar radius);
neutrino cooling only affects the thermal processes participating in the
mechanical structure.  Even at high luminosity, however, different neutrino 
rates result in only small changes in measurable quantities such as surface 
gravity.  Current observational techniques could not hope to resolve such
small differences.

\begin{figure}[h]
\includegraphics[scale=0.29,angle=-90]{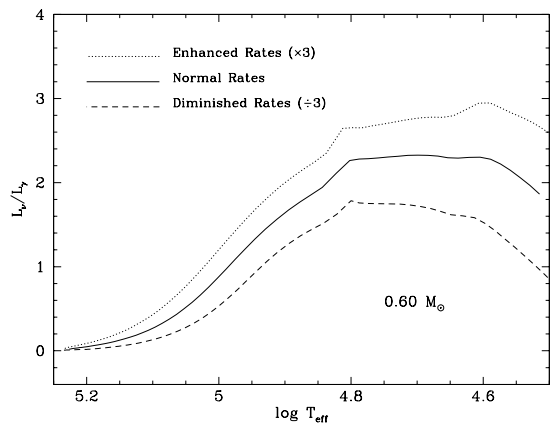}
\caption{Ratio of the neutrino luminosity to the photon luminosity as
a function of~$T_{\rm eff}$, for three~$0.6\,M_{\odot}$~model sequences
with different input neutrino production rates.}
\end{figure}

Figure~5 shows a more tangible effect of changing the rates.  Even
though models with different rates look much the same at a
given $T_{\rm eff}$, they get there at widely differing times, since
the rate of evolution along a track is directly dependent on
the importance of neutrino emission as a source of cooling.
For example, the model with enhanced neutrino rates cools from
$100,000$~K down to $65,000$~K in $600,000$~years, while the
model with diminished neutrino rates takes $1.3$~million years,
more than twice as long, to cool by the same amount.  The
maximum difference in the slope of the different curves in Figure~5 
occurs at $T_{\rm eff} \sim 80,000$~K.  Thus
the epoch where the rate of evolution is most sensitive to the
assumed neutrino physics corresponds to the position in the
H-R diagram occupied by the coolest pulsators in the PWD
instability strip.  On the other hand, for stars in the strip
hotter than $100,000$~K Figure~5 shows that
evolutionary rates do not depend on neutrino rates.

\begin{figure}[h]
\includegraphics[scale=0.29,angle=-90]{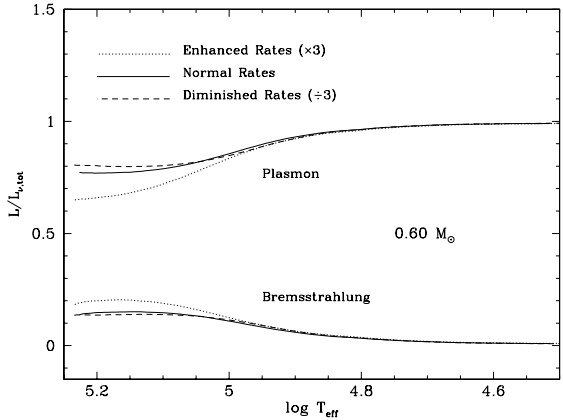}
\caption{Fraction of total neutrino luminosity contributed by
the plasmon and bremsstrahlung processes, as a function
of~$T_{\rm eff}$, for three~$0.6\,M_{\odot}$~model sequences
with different neutrino production rates.}
\end{figure}

Our expectations are borne out in Figure~6, which shows the rate of
change in period, $\dot{\Pi}/\Pi$ ($\equiv d(\ln\Pi)/dt$), as a function 
of period, $\Pi$, for PWD models at $140,000$~K (lower panel) 
and $80,000$~K (upper panel), given normal, enhanced, and diminished
neutrino production rates.  The rate of period
change $\dot{\Pi}/\Pi$ in the cooler models changes by a factor
of four between the enhanced and diminished rates.  Changing the
neutrino rates has little effect on $\dot{\Pi}/\Pi$ in the hotter
model, consistent with the results from Figure~5.  We now turn to
the exciting implications of these results, and explore the possibility 
and practicality of measuring $\dot{\Pi}/\Pi$ in cool pulsating 
PWD stars.  We can then identify likely targets for future 
observational campaigns.

\section{Prospects for Measuring Neutrino Cooling Effects}

\subsection{Determination of d$\Pi$/dt}

Unfortunately, the period changes expected to occur in PWD stars are too small 
to detect from simple comparison of the period from one year to that of the next.
To determine d$\Pi$/dt, a better technique is
to measure the cumulative phase change in a mode with a slowly changing period.
This is accomplished by comparing the observed times of maxima ($O$) in
the light curve to the times of maxima ($C$) calculated from an
assumption of constant period.  The resulting plot of ($O-C$) shows the
phase drift associated with a changing period.  A constant rate of period
change, d$\Pi$/dt, enters as a quadratic term in time:
\begin{equation}
(O-C) \approx \frac{1}{2} \frac{1}{\Pi_{t_{o}}} \frac{d\Pi}{dt}
(t-t_{o})^{2} \;\;\;{\rm [sec]}
\end{equation}
where $\Pi_{t_{o}}$ is the period at time $t_{o}$ (see for example Winget
et al.~1985, 1991 and Kepler et al.~1995).  To measure d$\Pi$/dt
with confidence, the star must of course have stable and fully resolved
pulsation periods, with reliable phase measurements from season to season.

\begin{figure}[h]
\plotone{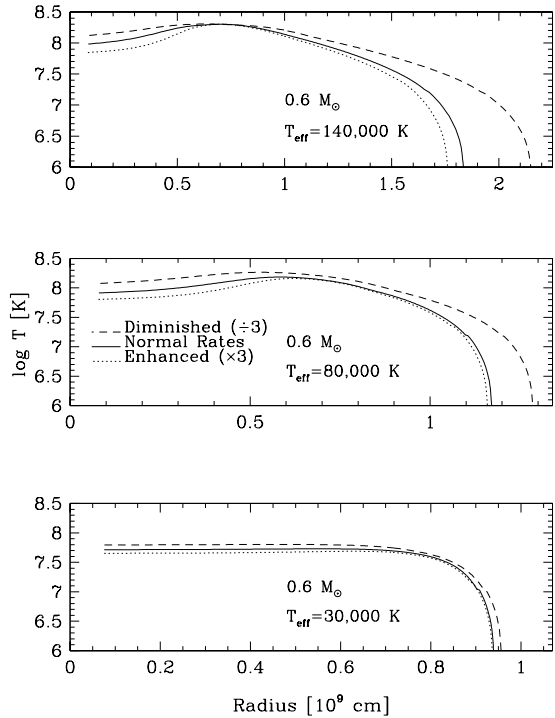}
\caption{Thermal structure at three different evolutionary
stages, $T_{\rm eff}=140,000$~K (upper
panel), $80,000$~K (middle panel), and $30,000$~K (lower
panel), for $0.6~M_{\odot}$ models with different neutrino
production rates.}
\end{figure}

\begin{figure}[h]
\includegraphics[scale=0.29,angle=-90]{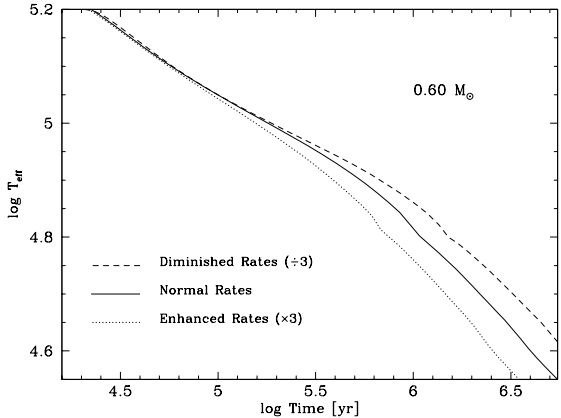}
\caption{Evolution of~$T_{\rm eff}$~with time for
three~$0.6\,M_{\odot}$~model sequences with different input neutrino
production rates.}
\end{figure}

Kawaler, Hansen, \& Winget~(1985) and Kawaler \& Bradley (1994) present predicted
values of d$\Pi$/dt for models relevant to GW~Vir and PNNV stars; the
only observed value of d$\Pi$/dt, that for PG~1159 itself, is consistent
with these models.  However, as Kawaler \& Bradley (1994) demonstrated,
for a star as hot as PG~1159, d$\Pi$/dt is strongly affected by mode
trapping.  This is an effect whereby some modes become excluded from
regions below subsurface composition discontinuities.  Kawaler \& Bradley (1994)
show that, in general, d$\Pi$/dt should be positive; this reflects the overall 
cooling of the model (Winget, Hansen, \& Van Horn~1983).  Trapped modes, however, are 
concentrated in the outer layers, within which contraction dominates cooling; 
therefore trapped modes can show periods which decrease with time.  Thus, mode 
trapping can complicate the interpretation of measured period changes in hot PWDs.
As GW~Vir stars cool, the surface contraction rate decreases
relative to the cooling rate of the interior.  So, while mode trapping
can still influence the pulsation period distribution itself, the rates of
period change become more similar from mode to mode in cooler GW~Vir
stars.  Kawaler \& Bradley (1994) found that the sign of d$\Pi$/dt could
be different for different modes in hot GW~Vir models; by the time those
models evolve to the cool end of the strip, the period change rates are
all positive.

\subsection{Prospective Targets}

Measurements of secular period change, $\dot{\Pi}/\Pi$, in white dwarfs have
been attempted by a number of investigations, with either
measurements made or tight upper limits set for the GW~Vir star PG~1159
(Winget et al.~1985, 1991, Costa \& Kepler 1998) and G117-B15A (Kepler et 
al.~1995).  Unfortunately, neutrino cooling is not expected to be an important 
effect for either of these stars.  On the other hand PG~0122, with a \teff~of 80,000~K, 
occupies the stage in GW~Vir evolution most highly dominated by neutrino 
emission.  O'Brien et al.~(1998) show that PG~0122 is in addition the most 
massive GW~Vir star, which should enhance neutrino effects as well.  

\begin{figure}[h]
\includegraphics[scale=0.37]{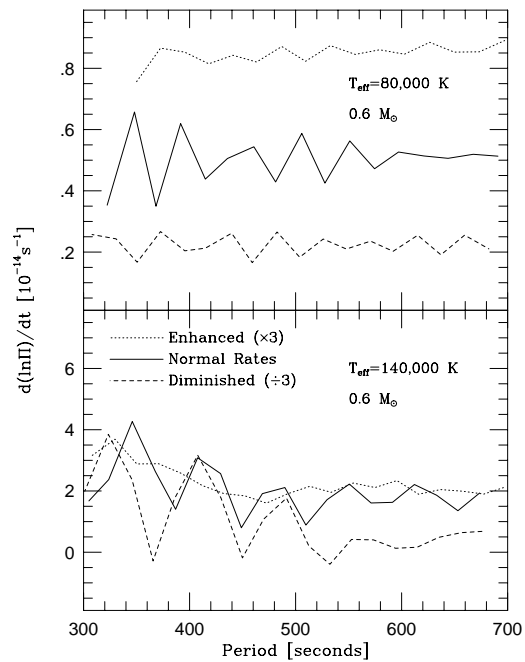}
\caption{Rate of period change for $0.6~M_{\odot}$ models with
different neutrino rates.}
\end{figure}

In order to measure $\dot{\Pi}/\Pi$ with confidence, a star must have a stable
and fully resolved pulsation, with stable phase measurements from season
to season.  PG~0122 is a very stable pulsator: over the past decade, it 
has shown a consistent pulsation spectrum, with the large-amplitude modes 
present at the same frequencies during each of three intensive observing 
seasons in 1986, 1990, and 1996.  The amplitudes of each of the dominant 
modes remained approximately constant as well (O'Brien et al.~1998).
Therefore, PG~0122 is an excellent candidate for measurement of the rate of 
secular period change caused by the evolutionary cooling of its interior.

In addition to the physics governing neutrino production, PG~0122 is
an ideal target for measuring neutrino emission rates because of the
minimal influence of any mode trapping on interpretation of its $\dot{\Pi}/\Pi$.
As mentioned above, for stars below 100,000~K trapping no longer significantly affects
$\dot{\Pi}/\Pi$ from mode to mode.

From Figure~6, we estimate the value of $\dot{\Pi}/\Pi$ for the dominant pulsation
mode ($\Pi = 400$~s) in PG~0122 to be about $6\times 10^{-15}$~sec~$^{-1}$.
With this rate of period change, the period should increase by about
$0.001$~s in 10 years; this is smaller than the period uncertainty
for a run length of several months (assuming a frequency precision of
$\frac{1}{10 \times {\rm run~length}}$).  However, the accumulated phase 
advance over a ten year period should be {\it nearly two full cycles}. 

Using the periods alone from the 1986, 1990, and 1996 data, O'Brien et al.~(1998) 
attempted to calculate $\dot{\Pi}/\Pi$ directly.  From the best least-squares periods
from 1996 and 1986, they calculate a period change of $-0.10 \pm 0.02$~s,
implying $\dot{\Pi}/\Pi = -7 \times 10^{-13}$~sec~$^{-1}$, about 100 times 
larger in magnitude---and different in sign---than theory expects.  However, 
as O'Brien et al.~(1998) point out, this calculation is based on the formal 
errors from a least-squares fit to the observed periods, and the 
{\it formal} least-squares error generally
underestimates the true error by an order of magnitude.  In practice, the
data currently available allow an upper limit to be set on the absolute
magnitude of $\dot{\Pi}/\Pi$
for PG~0122 of $1.5 \times 10^{-12}$~sec~$^{-1}$.  In view of the importance 
of measuring $\dot{\Pi}/\Pi$ for this star---as well as for the other cool 
GW~Vir stars---we must continue analysis of archival data and mount observing 
campaigns in the near future to monitor all the known GW~Vir stars with 
\teff~$< 100,000$~K.  With frequent observation, an accumulated phase advance 
of half a cycle, combined with the techniques described above, could be 
used to determine $\dot{\Pi}/\Pi$ for the GW~Vir stars PG~0122, PG~2131, and 
PG~1707 in two to three years.  In the case of PG~0122, the data presented in 
O'Brien et al.~(1998) provide a key anchor for this investigation.

\section{Summary and Conclusions}

We have shown that the predicted rates of period change in GW~Vir stars near
the cool end of the instability strip are sensitive to the neutrino production
rates used in stellar models.  The persistence of the solar neutrino problem 
has made the standard model of neutrino interactions one of the most intensely 
scrutinized theories in all of physics.  Determination of $\dot{\Pi}$ in 
the GW~Vir stars PG~0122 and PG~2131 will provide an important test of the 
standard model and of any new theories put forward to replace it.

\acknowledgments

The authors express their appreciation to Chris Clemens for valuable editorial 
comments.  We also thank the anonymous referee who, in particular, helped clarify 
our understanding and explanation of lepton scattering theory as it applies to 
white dwarf interiors.

MSO'B was supported during much of this research by a GAANN fellowship 
through grant P200A10522 from the Department of Education to Iowa State University.
Support also came from the National Science Foundation under the NSF Young Investigator 
Program (Grant AST-9257049) to SDK at Iowa State University.   Finally, some support 
for this work came to SDK from the NASA Astrophysics Theory Program through award 
NAG5-4060 to Iowa State University.

\end{document}